\documentclass[12pt]{iopart}

\usepackage{iopams}
\usepackage{graphicx}
\usepackage{color}
\begin{document}

\title[Dynamic Principles of COM]{Dynamic Principles of Center of Mass in Human Walking}

\author{Yifang Fan$^{1*}$, Mushtaq Loan$^{2*}$,
 Yubo Fan$^{3}$, Zhiyu Li$^{4}$ and Changsheng Lv$^{1}$}

\address{$^{1}$Center for Scientific Research, Guangzhou
Institute of Physical Education, Guangzhou 510500, P.R. China\\
$^{2}$International School, Jinan University, Guangzhou 510632, P.R. China\\
$^{3}$Bioengineering Department, Beijing University of
Aeronautics and Astronautics, Beijing 100191, P.R. China\\
$^{4}$College of Foreign Languages, Jinan University, Guangzhou
510632, P.R. China} \ead{tfyf@gipe.edu.cn and mushe@jnu.edu.cn}

\begin{abstract}
We present results of an analytic and numerical calculation that
studies the relationship between the time of initial foot contact
and the ground reaction force of human gait and explores the
dynamic principle of center of mass. Assuming the ground reaction
force of both feet to be the same in the same phase of a stride
cycle, we establish the relationships between the time of initial foot contact and the ground reaction
force, acceleration, velocity, displacement and average kinetic energy of center of
mass. We employ the
dispersion to analyze the effect of the time of the initial foot
contact that imposes upon these physical quantities. Our study
reveals that when the time of one foot's initial contact falls
right in the middle of the other foot's stride cycle, these
physical quantities reach extrema. An action function has been
identified as the dispersion of the physical quantities and
optimized analysis used to prove the least-action principle in
gait. In addition to being very significant to the research
domains such as clinical diagnosis, biped robot's gait control,
the exploration of this principle can simplify our understanding
of the basic properties of gait.
\end{abstract}

\maketitle

\section{Introduction}

Gait analysis plays an important role in exploring laws of human
motion by gait parameters via biomechanical methods. Many studies
have shown that gait parameters are significantly symmetric
\cite{Mitoma,Kim} and can be understood in terms of segmental
kinematical and kinetic physical quantities, while walking
\cite{Vaughan,Yang2001}. It has been shown,
\cite{Kondraske,Titianova,Liu2000,Biswasa} that by developing gait
parameters into evaluation indexes, one can assess the causal
relationship between physical injury and gait ability, which in
turn has been applied in rehabilitation therapy with success
\cite{Wall,Titianova,Liu1996,Sadeghi,Wang,Yang2005}.

The essence of human center of mass (COM) motion is actually the
periodical change which is under the effect of external forces
such as, ground reaction force (GRF), gravity and resistance
\cite{Kokshenev}. Previous studies indicate that the law of COM
motion is a more reliable gait evaluation method \cite{Elena}. The
common features of the frequently-used gait indexes lie in the
fact that normal human gait parameters have been regarded as the
criteria to evaluate rehabilitation. However, the incomplete
symmetry of human shape, coordinate and strength has formed the
uniqueness of human gait \cite{Murray}. This has brought
difficulty to the establishment of standard gait parameter index.
Consequently, exploring a principle that is relevant to gait
parameters such as cycle time, cadence or stride length in a
normal human gait has become essential to the study of gait
biomechanics \cite{Fan}. There are a number of issues we wish to
explore in our calculation. An important question is whether the
relationships between the GRF, velocity of COM, average kinetic
energy and the time of initial foot contact (tIFC) in a stride
cycle can be established. In particular we wish to examine if and
to what extent such relationships explore the principle behind the
gait characteristics. In the present work, we wish to advance an
attempt towards an approach where the precision of the calculation
reaches new levels.

The discovery of many basic principles originated from the study
of animal movements \cite{Kokshenev}. By examining the GRF acted
on the normal human gait (bare-footed), we establish the
relationships between the GRF, velocity of COM, average kinetic
energy and the tIFC in a stride
cycle. In this contribution we present our recent results on
kinetic regularities of COM. Another crucial issue in this
attempt concerns the prediction of action in gait. Using analytic
and numerical techniques, we shall identify the action in gait and
explore the possibility of the least-action principle in gait. The
rest of the paper is organized as follows:

In Sec. II we describe our procedure to establish the
relationships between the GRF and the tIFC and discuss the aspects
of determination of the force distribution in anteroposterior,
mediolateral and longitudinal directions. We address the problem
of the force distribution in transverse, sagittal and frontal
planes and optimization of dispersion of force. We present a
description of the determination of working parameters, such as
acceleration, velocity and displacement of COM and establish the
relationships between the kinetic regularities and the tIFC in
this section. We conclude this section by addressing the effect of
the tIFC upon the average kinetic energy of COM. Our main results
are presented and discussed in Sec. III. Finally, we present our
conclusions in Sec. IV.

\section{Analysis Strategy - Forward Dynamic Method}

\subsection{The Effect of tIFC upon GRF}
The analysis of COM's dynamic characteristics to evaluate the gait
features is rather non-trivial. The inverse dynamic method
\cite{Steven,Elena,Winiarski} has given mixed results with
systematic and statistical errors as major sources of
uncertainties in the analysis of COM dynamics characteristics
\cite{Fan2007,Ren}. An attempt to indicate the dynamics
characteristics of the whole body COM by one particular segment of
body will highly underestimate the results. The target of this
work is a calculation that gives more precise estimates of
dynamical measurements in question. We use forward dynamics method
\cite{Anderson,Chau} to illustrate on how tIFC determines the
force distribution and establish a relationship between the tIFC
and its force dispersion.

Since dynamic characteristics of COM in gait is governed by
kinematics and kinetics of COM, therefore, for constant gravity, the
analysis of force upon COM will compliment the analysis of GRF.
Also, GRF is caused by the body segmental movement driven by the
transarticular muscles and eventually by the contact of the foot
to the ground, so one foot's GRF includes the longitudinal GRF,
frontal friction and sagittal friction, i.e.,
\begin{eqnarray}
F(t)=F_{x}(t)\textbf{i}+F_{y}(t)\textbf{j}+F_{z}(t)\textbf{k},
\label{eqn3-1}
\end{eqnarray}
where $F(t)$ is the GRF of one moment at a stride cycle and
$F_{x}(t)$, $F_{y}(t)$ and $F_{z}(t)$ represent the
components in three directions, respectively. If $F_{i}^{l}(t)$
and $F_{i}^{r}(t)$ ($i=x,y,z$) represent both feet's sagittal and
frontal frictions and longitudinal GRF, respectively, then the
above equation shows a similarity of both feet's GRF distribution
when expressed in biped gait features \cite{Fan}. Consequently, we
assume that in each stride cycle, the left and right foot have an
identical distribution of GRF in all three directions. This would
imply that the GRF variations are only caused by the tIFC of both
feet. Setting $T$ as one foot's stride cycle time, the initial
phase of one foot is equal to zero and that of the other foot $t_{o}$
(i.e. tIFC), then corresponding to one foot's GRFs (for example left foot)
\begin{displaymath}
F^{l}_{x}(t)=F_{x}(t),\hspace{0.30cm} F^{l}_{y}(t)=F_{y}(t),
\hspace{0.30cm} F^{l}_{z}(t)=F_{z}(t),
\end{displaymath}
those of the other foot are
\begin{displaymath}
F^{r}_{x}(t)=F_{x}(t+t_{o}), \hspace{0.30cm}
F^{r}_{y}(t)=F_{y}(t+t_{o}), \hspace{0.30cm}
F^{r}_{z}(t)=F_{z}(t+t_{o})
\end{displaymath}
and Eq. (\ref{eqn3-1}) becomes
\begin{equation}
F(t,t_{o})=F_{x}(t,t_{o})\textbf{i}+F_{y}(t,t_{o})\textbf{j}+F_{z}(t,t_{o})\textbf{k}.
\label{eqn3-2}
\end{equation}
Since gait is a continuous and periodic movement, therefore while
walking at steady speeds, $F(t,t_{o})$ is the GRF when tIFC is $t_{o}$ and the stride cycle time is $t$. If $F(t,t_{o})=F(t+nT,t_{o}+nT)$
$(n=1,2,3\ldots)$ holds, then the GRF and gravity $(W)$ in a
stride cycle that act on the impulse of COM will follow the
following equation:
\begin{equation}
I=I_{x}\textbf{i}+I_{y}\textbf{j}+I_{z}\textbf{k}\equiv0.
\label{eqn3-3}
\end{equation}
To calculate the impulse characteristics of each foot's GRF in
gait when both feet have same GRF, we analyze two phases of one
foot's stride cycle. Let $T_{s}$ and $T_{w}$ represent the stance
phase and swing phase, respectively, then in the anterposterior
direction, Eqs. (\ref{eqn3-2}) and (\ref{eqn3-3}) yield
\begin{displaymath}
\int_{0}^{T} F_{x}(t)dt = 0
\hspace{0.40cm}\mbox{and}\hspace{0.40cm} \int_{0}^{T} F_{x}(t)dt =
\int_{0}^{T} F_{x}(t+t_{0})dt.
\end{displaymath}
Since $T=T_{s}+T_{w}$ and $\int_{T_{s}}^{T_{s}+T_{w}}F_{x}(t)dt
=0$, therefore
\begin{displaymath}
\int_{0}^{T_{s}} F_{x}(t)dt= \int_{0}^{T_{s}} F_{x}(t+t_{0})dt =0,
\end{displaymath}
whereas in the mediplateral direction, we get
\begin{displaymath}
\int_{0}^{T_{s}} F_{y}(t)dt= \int_{0}^{T_{s}} F_{y}(t+t_{0})dt.
\end{displaymath}
On the other hand, the calculated contribution in the
longitudinal direction gives
\begin{displaymath}
\int_{0}^{T} F_{z}(t)dt= \int_{0}^{T} F_{z}(t+t_{0})dt=
\frac{1}{2}\int_{0}^{T}W dt.
\end{displaymath}
The above equations give the impulse characteristics of each
foot's GRF in gait.
\subsection{Dispersion of GRF}
To analyze the dynamics that emerge in the GRF distribution at
$t_{0}=0, T/2$ and $T$, we bring forward the concept of dispersion
of GRF. If $\sigma_{i}$ $(i=x,y,z)$ and $\bar{F}_{i}$ denote the
GRF dispersions and average values of GRF in three direction, then
the corresponding correlation between the force dispersion and
average GRF can be written as
\begin{equation}
\label{eqn3-4} \sigma_{i}(t_{o})
=\sqrt{\sum^{T}_{t=0}\big(F_{i}(t,t_{o})-\bar{F}_{i}(t_{o})\big)^{2}\big/T\sum^{T}_{t=0}1},
\end{equation}
where $F_{i}(t,t_{0})$ is the same as that in Eq . (\ref{eqn3-2}) and $\sum^{T}_{t=0}1$ is the number of $\bigtriangleup t$ within the range $[0,T]$. For sequential
values of $t_{0}$ within the range $[0,T]$, we calculate the
values of $\sigma_{i}$ and use them to evaluate the effect of
tIFC upon GRF, velocity, position and kinetic energy of COM. We
use the average value of GRF from $20$ subjects to check the
reliability and accuracy of this method. This signature is
confirmed in Sec. III of this study.

\subsection{Regularities of velocity and displacement of COM}
Ground reaction force\footnote{We assume a constant gravity and a
negligible air resistance.} is the result of body segments action
on the ground via foot and determines the kinetic regularities
such as acceleration, velocity and displacement of human body COM.
Knowing the GRF and weight, the acceleration of COM at a given
instant in a stride cycle using Eq. (\ref{eqn3-2}) can be written
as
\begin{eqnarray}
a(t,t_{o})=F(t,t_{o})-W\textbf{k}.
\label{eqn4-1}
\end{eqnarray}
Expressing the acceleration of COM in component form, Eqs.
(\ref{eqn3-1}) and (\ref{eqn4-1}) reveal that the tIFC and GRF
have the same effect on the acceleration of COM.

The absolute motion of human COM relative to absolute inertia
reference frame is composed of convected motion and relative
motion. Following Kokshenev \cite{Kokshenev}, we define gait in
accordance with motion in Eq. (\ref{eqn3-3}), as ``walking at
steady speeds", thus we regard the convected motion a constant
parameter. Using Eq. (\ref{eqn4-1}) at tIFC = $t_{o}$, the COM
velocity in a stride cycle is given by
\begin{eqnarray}
v(t,t_{o})=\int^{t}_{0}a(t,t_{o})dt+v_{0}(t_{o}),
 \label{eqn4-2}
\end{eqnarray}
where $v_{0}(t_{o})$ is the initial velocity of COM in relative
motion at the beginning of a stride cycle. It seems that we cannot
confirm the magnitude of $v(t,t_{o})$ in Eq. (\ref{eqn4-2}) by
kinetic method (since everyone's gait speed is different).
However, the distinct feature is that once GRF and tIFC are
determined and when it observes the motion in Eq. (\ref{eqn3-3}),
$v_{0}(t_{o})$ must have a unique solution. We can then establish
the relationship between the initial velocity and acceleration in
relative motion in a stride cycle
\begin{eqnarray}
v_{0}(t_{o})=-\sum^{T}_{\lambda=1}\int^{\lambda}_{0}a(t,t_{o})dt.
\label{eqn4-3}
\end{eqnarray}
Eq. (\ref{eqn4-2}) reveals that $v(T,t_{o})=v(t_{o})$, which
implies that the cycle of velocity of COM is in accordance of
$v(T+nT,t_{o}+nT) (n=1,2,\cdots)$, that is to say, the end of one
stride cycle marks the beginning of the next stride cycle. This
explains the periodical characteristics of velocity of COM in
gait. What needs to be further illustrated is that the initial
velocity calculated using Eq. (\ref{eqn4-3}) refers that at the
beginning of a stride cycle, the body is in the state of steady
speeds. From Eqs. (\ref{eqn4-1}) and (\ref{eqn4-3}) we conclude
that tIFC has determined the initial velocity in relative motion, which has nothing to do the convected velocity.

Just like the velocity of COM, displacement of COM also involves
absolute, convected and relative displacements. We define the COM
initial displacement in relative motion of different tIFC as
$s_{0}(t_{o})$. Using Eq. (\ref{eqn4-2}), the displacement of COM
in relative motion at any moment has the following form:
\begin{eqnarray}
s(t,t_{o})=\int^{t}_{0}v(t,t_{o})dt+s_{0}(t_{o}) \label{eqn4-4}
\end{eqnarray}
and the correlation between the initial displacement of COM and
velocity of COM in relative motion in a stride cycle can be
written as
\begin{eqnarray}
s_{0}(t_{o})=-\sum^{T}_{\lambda=1}\int^{\lambda}_{0}v(t,t_{o})dt.
\label{eqn4-5}
\end{eqnarray}
Similar to the behaviour of the initial velocity in relative
motion, we verified that the initial displacement seems to be
independent of gait velocity, cadence and stride length.

\subsection{Average kinetic energy}
Nature has always minimized certain important quantities when a
physical process takes place \cite{Marion}. Bipedal walking has
enabled the continuous evolution of human gait \cite{Jenkins} and
eventually it has brought about the optimized gait \cite{Srinivasan}
and thus became a behavioral trait. In this behavior, the
force, acceleration and velocity all have their minimal values
when $t_{o}=\frac{1}{2}T$. To understand the physical
significance of these gait dynamic characteristics we explore the
issue of energy consumption. In gait, human segment movement is a
combination of agonist, antagonist and synergist. The movements
such as stretch or flexure all consume mechanical energy. On the
other hand, $E_{k}\geq 0$ in COM kinetic energy,
$\sum_{1}^{T}E_{k}(t)> 0$ in total kinetic energy and
$\bar{E}_{k}(t)>0$ in average kinetic energy in a stride cycle,
whereas the corresponding potential energy counterparts are zero
respectively. Using the COM kinetic energy in relative motion,
the description of mechanical energy consumption in gait
simplifies to
\begin{eqnarray}
E(t,t_{o})=\frac{1}{2}v^{2}(t,t_{o}), \label{eqn5-1}
\end{eqnarray}
and can be easily resolved in component form. Knowing $E_{i}$, we
can set up the relationships between tIFC and COM kinetic energy
by defining $t$ and $t_{o}$ in the interval $[0,T]$.

\subsection{Experimental Details}

The experimental measurements were carried on the following set of
equipments: Simi Motion 7.0 Three-Dimensional Movement Analysis
System; three Kistler $40\times60$ $(cm^{2})$ force plates; force
plate frequency: $2000Hz$, with a systematic uncertainty of $\pm
1\%$. The assembled force plate position has been fixed by
gradienter to ensure the force plates are on the same plane.
The measurements are taken at the sample frequency of $1000Hz$.
Before each measurement, the equipment is examined and returned to
zero-position.

Twenty female subjects with mean age $20.63\pm 0.76$ years, mean
height $163.12\pm3.72$ cm and mean weight $45.74\pm 3.30$ kg
participated in the study. A few trials of each level gait item were
administered to subjects as they ambulated on instrumented
positions. All the patients agreed to participate in the research,
and signed freely an informed consent form and study was carried
out according to the existing rules and regulations of our
institute's Ethnic Committee. Before the test, all the subjects
were thoroughly briefed about the procedures and matters needing
attention so that they all understand the purpose and the
requirements of the test. Each subject's medical history is
inquired so as to exclude subjects with diseases such as
pathological change, deformity or injury to make sure that their
physical conditions would meet the requirements of the test. When
measuring their gaits, we start from the subjects' standing
position, bare-footed (both feet disinfected by $75\%$ of
ethanol). A pre-test is to guarantee that after they walked three
steps, the subjects step on the platform and to make sure that
each force plate can measure one foot's stance phase data
separately. When the subject's gait is found to be quite abnormal,
for example, it is obviously discontinuous, she would be asked to
perform again so that the recorded data meet the requirements of
the test.

Based on the fact that the longitudinal GRF in gait is apparently
greater than the GRFs of the anteroposterior or mediolateral
direction, we take the signal of longitudinal GRF to identify the
instants of initial foot contact and of terminal stance. The gait
we study is walking at steady speeds; therefore, we de-noise the
signals of the platform and examine the test signals would meet
the requirements in the predetermined accuracy range. We rate the
gait cycle time by percentage, normalize the weight and
standardize the GRF from three directions. We count on average of
the processed data of 20 subjects' three-direction GRF from
individual stride cycle to conduct our research.

\section{Results and Discussion}
\subsection{Spatial GRF, COM velocity and displacement}
\begin{figure}[!ht]
\begin{center}
\begin{tabular}{cccc}
 \includegraphics[width=13.8cm]{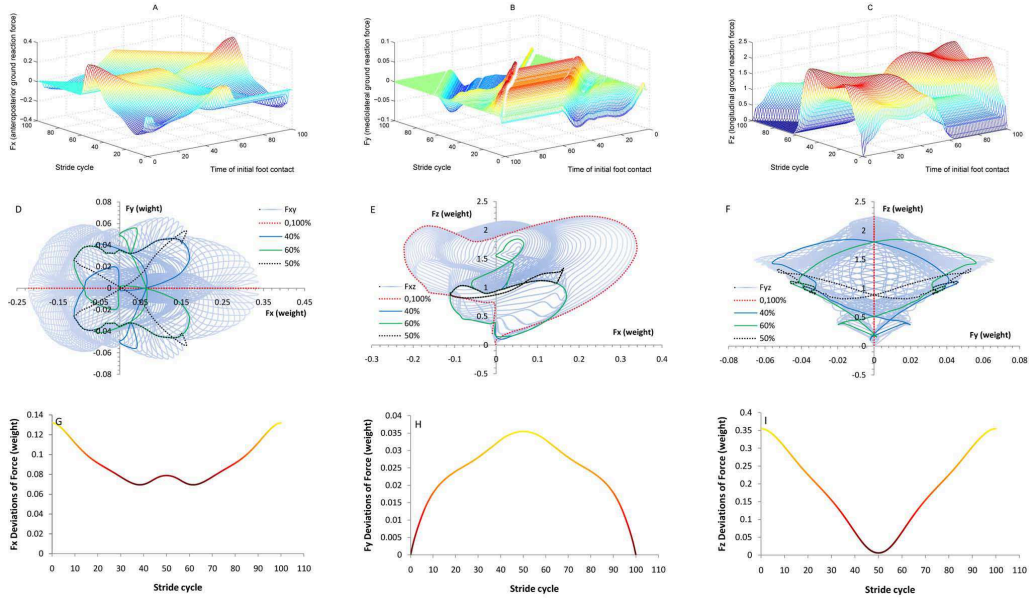}
\end{tabular}
\caption{\label{fig1} Schematic illustration of the observed
spectrum for tIFC and GRF showing the relationship between tIFC
and GRF in three directions (Fig.1A - C), together with tIFC and
GRF resultant force in three planes (Fig.1D - F), compared with
a typical tIFC and dispersion of force correspondence (Fig.1G - I).}
\end{center}
\end{figure}

To explore the spatial GRF, we examine Eq. (\ref{eqn3-2}) by using
the GRF numeric of $20$ subjects. Fig. \ref{fig1} collects and
displays the results of anterposterior, mediolateral and
longitudinal GRF. The GRF has been standardized ($F_{i}(t)/W)$)
and stride cycle time is rated by percentage. Fig. \ref{fig1}A - C
shows that tIFC has an effect upon the GRF in three directions.
Using three sets of component data from GRF of the subject's gait,
we obtained curved surface effective plots and the quantitative
relationship between tIFC and GRF resultant forces on three
planes.
Fig. \ref{fig1}D - F indicates that no matter what changes $t_{o}$
undergoes, GRF resultant forces in these three planes remain to be
closed curves. This significantly compliments and confirms the
dynamics behind Eq. (\ref{eqn3-3}), i.e., when the distribution
of one foot's GRF is determined and both feet's GRFs remain the
same, the momentum of COM always remains unaffected.

Fig. \ref{fig1}D - F also reveals the geometrical characteristics of
the plane resultant force. On the transverse plane, when $t_{o}=0,T$, the gait becomes a jump and its frontal resultant force
becomes a straight line. The resultant force changes into a
symmetrical butterfly about $F_{x}$, which could imply a
possible transformation at $t_{o}=\frac{1}{2}T$. On the sagittal
plane, when $t_{o}=0,T$, the resultant force
$F_{xz}$ forms the longest closed curve (the length of which
is calculated by $\oint F_{xz}(t,t_{o})dF_{x}dF_{y}$) and more
or less shortest at $t_{o}=\frac{1}{2}T$. On the frontal plane,
when the resultant force develops into a straight line at
$t_{o}=0,T$, the resultant force becomes a symmetrical
butterfly about $F_{z}$ at $t_{o}=\frac{1}{2}T$.

Having developed the relationship between tIFC and GRF's resultant
force on three planes, we calculate the GRF dispersion in three
directions using Eq. (\ref{eqn3-4}). To confirm the results beyond
doubt, we identify the GRF dispersion by analyzing the effective
plots shown in Fig. \ref{fig1}G - I. It is clear from
the plots that at $t_{o}=\frac{1}{2}T$, $\sigma_{x}$ has its
maximum value at the bottom region of "W" shape, $\sigma_{y}$ has
a global maxima and $\sigma_{z}$ a global minima. Using the
numerical values, $min(\sigma_{x})=0.0695$,
$max(\sigma_{x})=0.1319$, $min(\sigma_{y})=0.0000$,
$max(\sigma_{y})=0.0355$, $min(\sigma_{z})=0.1086$ and
$max(\sigma_{z})=0.9144$, of GRF dispersion obtained in this case
study, we find that the largest differences between the maximal
and minimal values are $0.0624$, $0.0355$ and $0.8058$,
respectively. This means that the dominant effect to the global
GRF is the longitudinal GRF. The minimal GRF dispersion on
longitudinal direction occurs at $t_{o}=\frac{1}{2}T$. The
analysis of the distribution of GRF in three directions indicates
that when $t_{o}=\frac{1}{2}T$, the anteroposterior and
longitudinal dispersions of GRF are almost the least and the
mediolateral one is the largest. The closed curve shaped by the
two components on the sagittal plane seems to be the shortest. The
GRF resultant force on the transverse and sagittal plane is shown
as symmetric butterfly.

\begin{figure}[!ht]
\begin{center}
\begin{tabular}{cccc}
 \includegraphics[width=13.8cm]{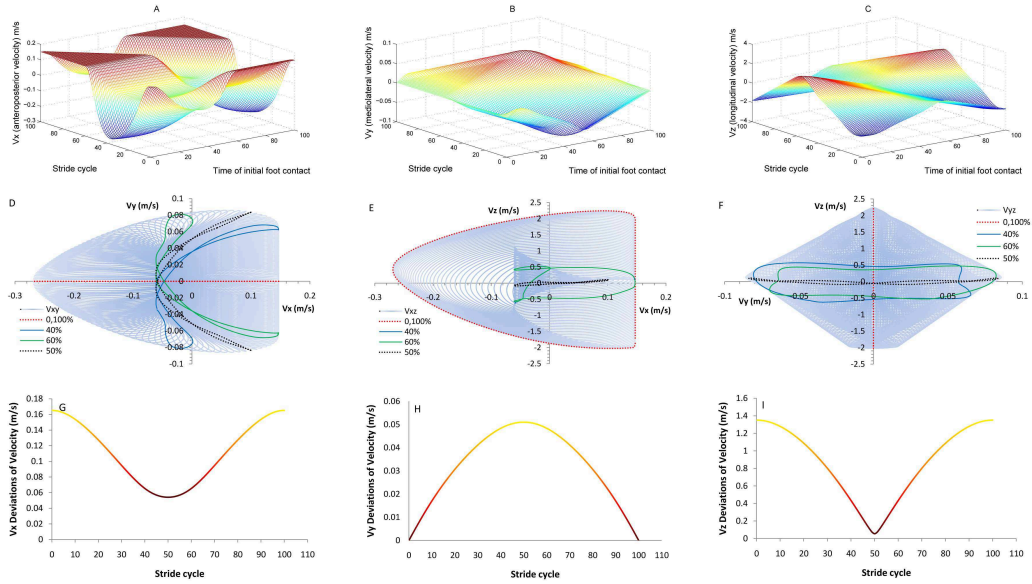}
\end{tabular}
\caption{\label{fig2} Effective plots showing relationship between
tIFC and COM velocity in three directions (Fig.2A - C), tIFC and
COM velocity in three planes (Fig.2D - F) and tIFC and
dispersion of COM velocity (Fig.2G - I).}
\end{center}
\end{figure}

Turning our attention towards the kinetic regularities, we plot
the effect of tIFC on COM velocity in Fig. \ref{fig2}. We can see
two platforms in Fig. \ref{fig2}A - C, which emerge at the two extremes
of tIFC rated by percentage forming a concave region around tIFC
whereas the velocity describes a convex region around tIFC. The
COM velocity in the longitudinal direction indicates that tIFC
entails the vertical velocity of COM to have a distribution of
tIFC in the form of a saddle. In order to further understand the
effect of tIFC exerting on the velocity of COM, we analyze the
variations of the velocity of COM in three planes.
Looking at Fig. \ref{fig2}D - F, we
notice that in the transverse plane, the direction of COM velocity
is a straight line at $t_{o}=0$ and $T$ and the plane velocity
becomes a symmetrical closed curve about $v_{x}$ axis at
$t_{o}=\frac{1}{2}T$. In the sagittal plane the closed curve
formed by the plane velocity is the longest (the length of the
closed curve is calculated by $\oint v(t,t_{o})dv_{x}dv_{z}$) at
$t_{o}=0,T$, and is shorter (not the shortest) at $t_{o}=0$
and $T$. In the frontal plane, when $t_{o}=0,T$, the
velocity becomes a straight line and the resultant velocity
becomes a symmetrical closed curve about $v_{z}$ axis at
$t_{o}=\frac{1}{2}T$. Having obtained the estimated value of COM
velocity, we are ready to set up the dispersion of relationship
between tIFC and COM velocity (as shown in Fig. \ref{fig2}G - I) using Eq. (\ref{eqn3-4}). We notice that at
$t_{o}=\frac{1}{2}T$, $\sigma_{x}$ and $\sigma_{y}$ have the
global minima and $\sigma_{z}$ has the global maxima. The
estimates, $min(\sigma_{x})=0.0541$, $max(\sigma_{x})=0.1652$,
$min(\sigma_{y})=0.0000$, $max(\sigma_{y})=0.0510$,
$min(\sigma_{z})=0.0549$ and $max(\sigma_{x})=1.3502$ in our case
study, result in largest differences of maximal and minimal values
of $0.1111$, $0.0510$, and $1.2953$ respectively. This means that
the dominant effect upon COM velocity lies in the longitudinal and
anteroposterior directions, where the dispersions have the minimal
values at $t_{o}=\frac{1}{2}T$.

\begin{figure}[!ht]
\begin{center}
\begin{tabular}{cccc}
 \includegraphics[width=13.8cm]{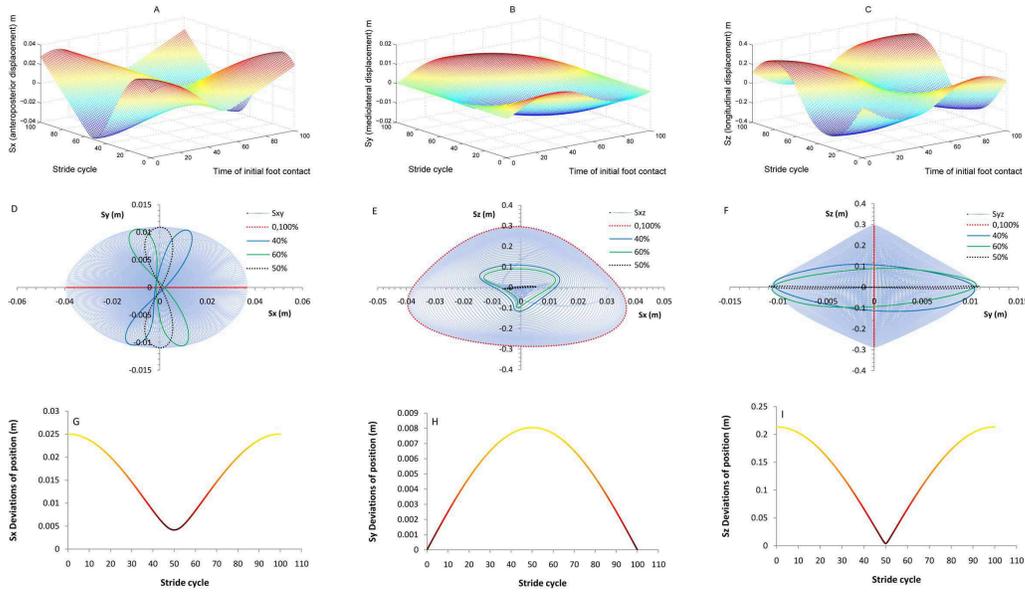}
\end{tabular}
\caption{\label{fig3} COM displacement as a function of tIFC. The
directional and planer components are shown in Fig.3A - C and Fig.3D - F, respectively. Fig.3G - I shows the dispersion of COM
displacement.}
\end{center}
\end{figure}

The kinematic regularity of the displacement of COM in three
directions is shown in Fig. \ref{fig3}A - C. We see
a geometric distribution of convex corners indicating that tIFC
changes the position of COM in anteroposterior and mediolateral direction. The
bimodal plot indicates the complexity of longitudinal direction
exerted by tIFC to the position of COM. The estimated
values of the COM displacement components are used to evaluate the
values and directions of COM displacement in three planes. To further analyze the
effect of tIFC upon the position of COM, we explore the changes of
COM positions from three planes which are displayed in Fig. \ref{fig3}D - F.

These effective plots show that in the transverse plane the COM
displacement is a straight line at $t_{o}=0,T$. On the
other hand a closed curve is formed by $s_{x}$ and $s_{y}$ and is
symmetric about $s_{x}$ axis at $t_{o}=\frac{1}{2}T$. On the
sagittal plane, when $t_{o}=0,T$, the closed curve
shaped by the sagittal displacement is the longest (the length of
the closed curve is calculated by $\oint s(t,t_{o})ds_{x}dy_{z}$);
when $t_{o}=\frac{1}{2}T$, the curve is rather shorter (but not
the shortest). On the frontal plane the frontal displacement
becomes a straight line at $t_{o}=0,T$ and a symmetric
closed curve about $s_{y}$ axis at $t_{o}=\frac{1}{2}T$.

Using the numerical values, $min(\sigma_{x})=0.0042$,
$max(\sigma_{x})=0.0252$, $min(\sigma_{y})=0.0000$,
$max(\sigma_{y})=0.0081$, $min(\sigma_{z})=0.0039$ and
$max(\sigma_{z})=0.2147$, obtained in this case study, we find
that of the GRF dispersion $\sigma_{x}$ and $\sigma_{y}$ have a
global minima and $\sigma_{z}$ has a global maxima at around
$t_{o}=\frac{1}{2}T$, again resulting in largest differences
between the minimal and maximal values with a magnitude of the
order $0.0210$, $0.0081$ and $0.2108$, respectively. This implies
that the greatest effect to COM displacement comes from components
on longitudinal and anteroposterior directions while the minimal
dispersion of displacement on the longitudinal and anteroposterior
direction emerges at $t_{o}=\frac{1}{2}T$ (See Fig. \ref{fig3}G - I).

\begin{figure}[!ht]
\begin{center}
\begin{tabular}{cccc}
 \includegraphics[width=13.8cm]{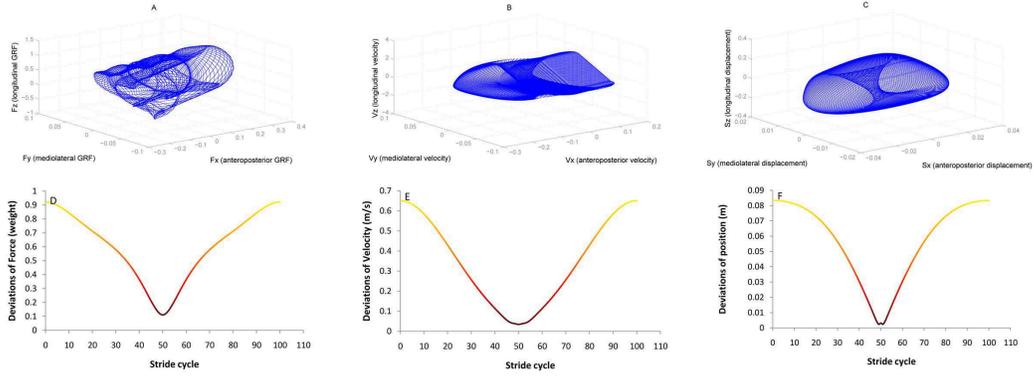}
\end{tabular}
\caption{\label{fig4} Relationship between tIFC and dispersion of
force.}
\end{center}
\end{figure}

Accordingly, at $t_{o}=\frac{1}{2}T$, COM velocity in relative
motion presents its symmetric closed curve in transverse and
frontal plane and the curve length in the sagittal plane is
approximately the minimal. In a stride cycle, the dispersion of
COM velocity in relative motion has the global minima on the
anteroposterior and longitudinal direction while the global
maximal value on the mediolateral direction. The effect of tIFC
upon the COM velocity and displacement does not depend on gait
velocity, cadence or stride length whereas the effect of the tIFC
upon COM acceleration is in accordance with the effect it exerts
upon the GRF. The COM acceleration and GRF show more or less
identical behaviour. Fig. \ref{fig4}A - C demonstrate the effect of tIFC upon three components of GRF, COM velocity and COM displacement on three direction. As is clear from Fig. \ref{fig4}D - F that at half stride cycle ($t=T/2$), the dispersion of
GRF and COM kinematics are minimum. In comparison to GRF and COM
velocity, the COM displacement shows a sharp dip with least
minimum.

\subsection{Anteroposterior, Mediolateral and longitudinal average
kinetic energies}
\begin{figure}[!ht]
\begin{center}
\begin{tabular}{cccc}
 \includegraphics[width=13.8cm]{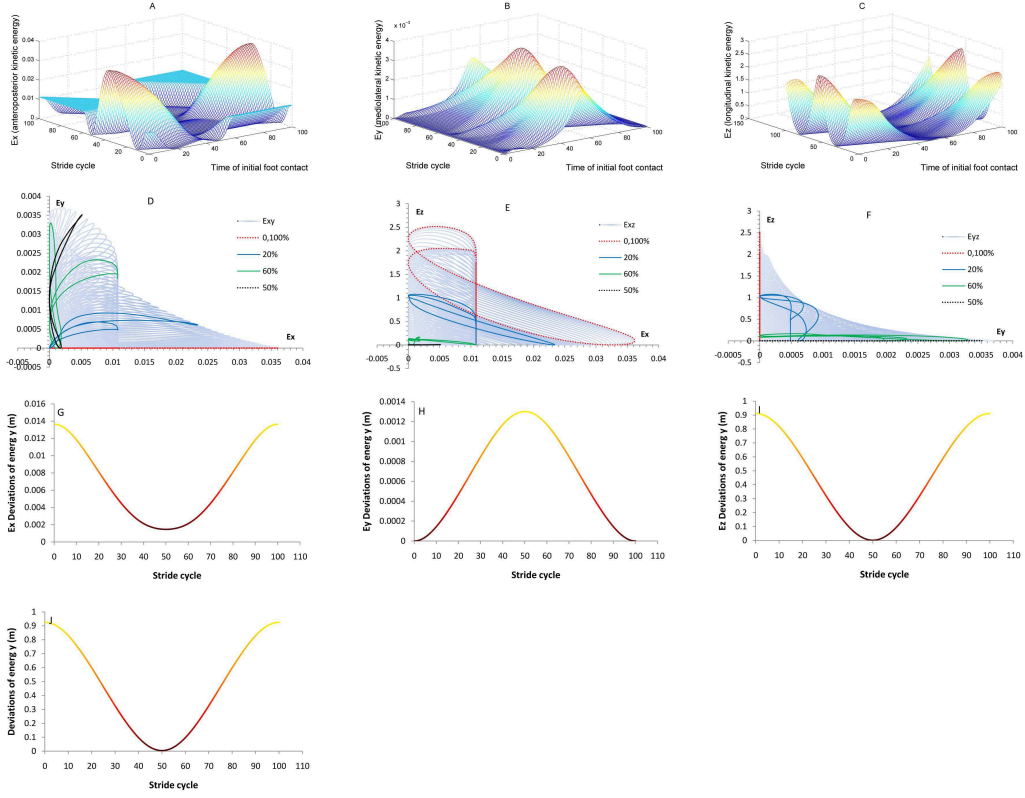}
\end{tabular}
\caption{\label{fig5} Relationship between tIFC and kinetic
energy. (Fig.5A - C tIFC and COM kinetic energy in three
directions; Fig.5D - F -tIFC and COM kinetic energy in three
planes; Fig.5G - I - tIFC and average kinetic energy. )}
\end{center}
\end{figure}

Finally, we show the COM kinetic energies
as a function of stride cycle in anteroposterior, mediolateral and
longitudinal channels in Fig. \ref{fig5}A - C, together
with corresponding transverse, sagittal and frontal energies in
three planes (Fig. \ref{fig5}D - F). Fig. \ref{fig5}G - I indicates that tIFC
makes COM average kinetic energy the global minima on
anteroposterior and mediolateral direction while that on the
mediolateral direction a global maxima at $t_{o}=\frac{1}{2}T$.
Comparing the relationship between the tIFC's velocity of COM and
the position of COM, tIFC contributes a symmetric distribution of
kinetic energy of COM. Since
$min(E_{x}(t,t_{o}))>max(E_{y}(t,t_{o}))$ and
$max(E_{y}(t,t_{o}))<min(E_{z}(t,t_{o}))$, the anteroposterior and
longitudinal average kinetic energies determine the COM average
kinetic energy.
when tIFC falls at the $20\%, 50\%, 60\%$ and $100\%$ of a stride
cycle, the kinetic energy of COM in three planes varies rather
considerably. On the transverse and frontal planes, the effect is
similar whereas the sagittal plane undergoes a non-trivial change.
We apply the dispersion of kinetic energy of COM to evaluate this
effect. We notice, from Fig. \ref{fig5}J, that COM average kinetic
energy is minimum at $t_{o}=\frac{1}{2}T$, which implies a minima
for the COM mechanical energy consumption and signatures the
physical significance of gait.

\subsection{Least-action principle in gait}
The intriguing question that arises from the above discussion if
there is a symmetry or a physical principle that is hidden in the
phenomenon that makes the COM kinetic energy to reach its
extremum at $t_{o}=\frac{1}{2}T$. In order to explore this issue,
we pursue the technique of optimizing the dispersion objective
function \cite{Fan}
\begin{eqnarray}
\sum^{T}_{1}(f(t)+f(t+t_{o})-\overline{f}(t_{o}))^{2}=\sum^{T}_{1}f(t)^{2}+\sum^{T}_{1}f(t+t_{o})^{2}\nonumber\\
+2\sum^{T}_{1}f(t)f(t+t_{o})-2\sum^{T}_{1}(f(t)+f(t+t_{o}))+(\overline{f}(t_{o}))^{2}\sum^{T}_{1}1\label{eqn5-3}
\label{eqn5-2}
\end{eqnarray}
The action of GRF and gravity are the reasons for human COM motion changes. Let's first analyze the GRF. The average value of the resultant force in the anteroposterior and mediolateral direction $\bar{F}_{x}(t_{o})$ and $\bar{F}_{y}(t_{o})$ is zero and in the longitudinal direction $\bar{F}_{z}(t_{o})$ is one (when the weight has been normalized). Therefore, for a known gait, $\sum^{T}_{1}(F_{x}(t))^{2}, \sum^{T}_{1}(F_{y}(t))^{2}, \sum^{T}_{1}(F_{z}(t))^{2}, \sum^{T}_{1}(F_{x}(t+t_{o}))^{2}, \sum^{T}_{1}(F_{y}(t+t_{o}))^{2}$ and $\sum^{T}_{1}(F_{z}(t+t_{o}))^{2}$ are all constants and Eq. (\ref{eqn5-2}) simplifies the dispersion optimization to
\begin{eqnarray}
\label{eqn5-4}
\min\psi_{F_{x}}(t_{0}) &=&
\sum_{1}^{T}F_{x}(t)F_{x}(t+t_{o})\nonumber\\
\max\psi_{F_{y}}(t_{0})&=&\sum_{1}^{T}
F_{y}(t)F_{y}(t+t_{o})\nonumber\\
\min\psi_{F_{z}}(t_{0})&=&\sum_{1}^{T}F_{z}(t)F_{z}(t+t_{o}).
\end{eqnarray}
To find a solution of the optimization problem, we need to set up
a function of GRF that changes with time in three directions. It
seems that the pulse periodic GRF fits the criteria and hence we
use segment trigonometric functions
\begin{displaymath}
F_{x}(t) = a\sin(2\pi\frac{t}{T_{s}}), \hspace{0.50cm}F_{y}(t) =
b\sin(\pi\frac{t}{T_{s}}) \hspace{0.50cm} F_{z}(t) =
c\sin(\pi\frac{t}{T_{s}}),
\end{displaymath}
where $a, b, c$ stand for the GRF after normalization in three directions and $T_{s}$ is the stance time. For a small enough test
frequency, the above equation transforms to
\begin{eqnarray}
\label{eqn5-5}
\min\psi_{F_{x}}(t_{0})&=&A\int^{T_{s}}_{0}\sin(2\pi\frac{t}{T_{s}})\sin(2\pi\frac{t+t_{o}}{T_{s}})dt\nonumber\\
\max\psi_{F_{y}}(t_{0})&=&-B\int^{T_{s}}_{0}\sin(\pi\frac{t}{T_{s}})\sin(\pi\frac{t+t_{o}}{T_{s}})dt\nonumber\\
\min\psi_{F_{z}}(t_{0})&=&C\int^{T_{s}}_{0}\sin(\pi\frac{t}{T_{s}})\sin(\pi\frac{t+t_{o}}{T_{s}})dt.
\end{eqnarray}
Let's take longitudinal GRF in Eq. (\ref{eqn5-5}) as an example.
In order to get the antiderivative of the integrand of integral
variable $t$, we detach $t$ and $t_{o}$ of trigonometric function
to obtain
\begin{eqnarray}\label{eqn5-6}
\int^{T_{s}}_{0}\sin(\frac{t}{T_{s}}\pi)\sin(\frac{t+t_{o}}{T_{s}}\pi)dt=\int^{T_{s}}_{0}(\sin^{2}(\frac{t\pi}{T_{s}})\cos(\frac{t\pi}{T_{s}})\nonumber\\
+\sin(\frac{t\pi}{T_{s}})\cos(\frac{t\pi}{T_{s}})\sin(\frac{t_{o}\pi}{T_{s}}))dt,
\end{eqnarray}
Now we can transform Eq. (\ref{eqn5-6}) into the integral of
integral variable $t$ while regarding the GRF as segment function.
The transformation reduces the pulse periodic GRF functions into
three piecewise functions at three intervals\footnote{While
walking $T_{s}+T_{w}=T$ and $T_{s}>T_{w}>0$.}: $[0,T_{w}]$,
$[T_{w},T_{s}]$ and $[T_{s},T]$ which, to the confirmed GRF, $a, b, c$ are constants and $A, B, C$ (related to $a, b, c$) are also constants, yields
the following form of longitudinal GRF:
\begin{equation}
\label{eqn5-7}
\min\psi_{F_{z}}(t_{o})=C\cases{\frac{1}{2}\sin(\pi\frac{t_{o}}{T_{s}})-(\frac{t_{o}-T_{s}}{2T_{s}}\pi)\cos(\pi\frac{t_{o}}{T_{s}})\nonumber\\
0\leq t_{o}\leq T_{w}\\
\frac{1}{2}\sin(\pi\frac{t_{o}}{T_{s}})-(\frac{t_{o}-T_{s}}{2T_{s}}\pi)\cos(\pi\frac{t_{o}}{T_{s}})\nonumber\\+\frac{1}{2}\sin(\frac{t_{o}-T_{w}}{T_{s}}\pi)
-(\frac{t_{o}-T_{w}}{2T_{s}}\pi)\cos(\frac{t_{o}-T_{w}}{T_{s}})\nonumber\\
T_{w}\leq t_{o}\leq T_{s}\\
\frac{1}{2}\sin(\frac{t_{o}-T_{w}}{T_{s}}\pi)-(\frac{t_{o}-T_{w}}{2T_{s}}\pi)\cos(\frac{t_{o}-T_{w}}{T_{s}}\pi)\nonumber\\
T_{s}\leq t_{o}\leq T\\
}
\end{equation}
The contribution
$(\frac{t_{o}-T_{s}}{2T_{s}}\pi)\sin(\pi\frac{t_{o}}{T_{s}})=0$ in
the interval $[0,T_{w}]$, thus giving the $t_{o}=0$, which in
turn yield (from Eq. (\ref{eqn5-7})) a maxima ($=\frac{C\pi}{2}$).
On the other hand, the contribution
$(\frac{t_{o}-T_{s}}{2T_{s}}\pi)\sin(\pi\frac{t_{o}}{T_{s}})+(\frac{t_{o}-
T_{w}}{2T_{s}}\pi)\sin(\pi\frac{t_{o}-T_{w}}{T_{s}})=0$ in
$[T_{w},T_{s}]$ giving the solution $T_{s}-t_{o}=t_{o}-T_{w}$
which yields
$C(\cos(\frac{T_{w}}{2T_{s}}\pi)-(\frac{T_{s}-T_{w}}{2T_{s}}\pi)\sin(\frac{T_{w}}{2T_{s}}\pi))$
(a minima). Similarly, we obtain a maximum value $\frac{C\pi}{2}$
in $[T_{s},T]$.

Following the above procedure, it is easy to verify (from Eq. \ref{eqn5-5}) that at $t_{o}=T/2$ the dispersion of GRF in the
anteroposterior direction attains a minima of
$A((\frac{T_{w}-T_{s}}{T_{s}}\pi)\cos(\frac{T_{w}}{T_{s}}\pi)-\sin(\frac{T_{w}}{T_{s}}\pi))$
and a maxima of
$B((\frac{T_{s}-T_{w}}{2T_{s}}\pi)\sin(\frac{T_{w}}{2T_{s}}\pi)-\cos(\frac{T_{w}}{2T_{s}}\pi))$.
in the mediolateral direction. Since, in gait, $A>B>C$, the sum
of dispersions in three directions is minimal at
$t_{o}=\frac{1}{2}T$. This confirms our results shown in Figs.
\ref{fig4}D - F and \ref{fig5}J. Similarly, the dispersion of COM
acceleration, dispersion of COM velocity and the COM mechanical
energy consumption are all the minimal at $t_{o}=\frac{1}{2}T$.
This phenomenon is independent of the physiological factors such
as height, weight and gait parameters \cite{Fan}.

\section{Conclusion}
In periodic motion, the foot's stance and swing substitute one
another such that one foot always remains in stance. This allows
the human body to be acted by the periodical GRF, which is related
not only to the foot's movement style, but also to the
substitution style of one foot with another. Following the biped
movement style in normal gaits and the similar traits of GRF, we
have been able to study the effect of tIFC upon the human COM
dynamic characteristics based on the assumption that both feet's
GRFs are the same in the same phase. Our results suggest that
when tIFC falls in the middle of the other foot's stride cycle,
the COM kinematics and dispersions of GRF acted on COM are the
minimal, which has entailed the minimal average MEC of muscles.
Our analysis suggests that it falls into the category of
least-action principle and is consistent with the phenomenon that
exists in normal gaits \cite{Fan}.

Based upon the least-action principle, we have observed that tIFC
has caused the GRF and the COM regularities to form a closed-curve
on the transverse and frontal plane, which might present a new
method for gait evaluation. We advocate the use of this method to
uncover and speed up the diagnosis simply by measuring the GRF
for the patients with foot injuries or arthritis (for such
patients, the model for these quantities is not symmetric
\cite{Kfc}). Meanwhile, the patients need only walk a few steps in
their normal gait, their COM dynamic characteristics will be
acquired easily and more accurately. This is exactly what the
clinic diagnosis is looking for. In addition to the human gait's
adaptation to natural environment \cite{Jenkins,Richmond}, the
evolution of gait is the result of its observation of
least-action principle. We believe that precise measurements of
the variations of shear stress in natural gaits (bare-footed)
shall profoundly enriched the content of least-action principle
\cite{Liu2007}. A further study of least-action principle will be
significant to the domains such as sport rehabilitation, biometric
identification \cite{Boulgouris,Nixon} and control of biped robot
gaits \cite{Collins2001,Collins,Ohgane}.

\section*{Acknowledgments}

This project was funded by National Natural Science Foundation of
China under the grant $10772053$, $10972061$ and by Key Project of
Natural Science Research of Guangdong Higher Education Grant No
$06Z019$. The authors would like to acknowledge the support from
the subjects.

\end{document}